\documentclass[12pt]{article}

\def\lsim{\mathrel{\rlap{\lower3pt\hbox{\hskip0pt$\sim$}}
     \raise1pt\hbox{$<$}}}         
\def\gsim{\mathrel{\rlap{\lower4pt\hbox{\hskip1pt$\sim$}}
     \raise1pt\hbox{$>$}}}         

\usepackage{amsmath}
\usepackage{amsfonts}

\begin{document}
\begin{titlepage}

\centerline{\Large \bf Non-perturbative Massive Solutions in}
\centerline{\Large \bf Gravitational Higgs Mechanism}
\medskip

\centerline{Zurab Kakushadze$^\S$$^\dag$\footnote{\tt Email: zura@quantigic.com}}

\bigskip

\centerline{\em $^\S$ Quantigic$^\circledR$ Solutions LLC}
\centerline{\em 1127 High Ridge Road \#135, Stamford, CT 06905\,\,\footnote{DISCLAIMER: This address is used by the corresponding author for no
purpose other than to indicate his professional affiliation as is customary in
scientific publications. In particular, the contents of this paper are limited
to Theoretical Physics, have no commercial or other such value,
are not intended as an investment, legal, tax or any other such advice,
and in no way represent views of Quantigic$^\circledR$ Solutions LLC,
the website {\underline{www.quantigic.com}} or any of their other affiliates.}}
\centerline{\em $^\dag$ Department of Physics, University of Connecticut}
\centerline{\em 1 University Place, Stamford, CT 06901}
\medskip
\centerline{(May 7, 2013)}

\bigskip
\medskip

\begin{abstract}
{}We construct exact non-perturbative massive solutions in the gravitational Higgs mechanism. They confirm the conclusions of arXiv:1102.4991, which are based on non-perturbative Hamiltonian analysis for the relevant metric degrees of freedom, that while perturbatively unitarity may not be evident, no negative norm state is present in the full nonlinear theory. The non-perturbative massive solutions do not appear to exhibit instabilities and describe vacuum configurations which are periodic in time, including purely longitudinal solutions with isotropic periodically expanding and contracting spatial dimensions, ``cosmological strings" with only one periodically expanding and contracting spatial dimension, and also purely non-longitudinal (``traceless") periodically expanding and contracting solutions with constant spatial volume. As an aside we also discuss massive solutions in New Massive Gravity. While such solutions are present in the linearized theory, we argue that already at the next-to-linear (quadratic) order in the equations of motion (and, more generally, for weak-field configurations) there are no massive solutions.

\end{abstract}
\end{titlepage}

\newpage

\section{Introduction and Summary}

{}The gravitational Higgs mechanism gives a non-perturbative and fully
covariant definition of massive gravity \cite{KL, thooft, ZK, Oda, ZK1, ZK2,
Oda1, Demir, Cham1, Oda2, JK, dS, Cham2, Ber2, Cham3, Unitarity}. The graviton acquires mass via spontaneous
breaking of the underlying general coordinate reparametrization invariance by
scalar vacuum expectation values.\footnote{For earlier and subsequent related
works, see, {\em e.g.}, \cite{Duff, OP, GMZ, Perc, GT, Siegel, Por, AGS,
Ch, Ban, AH1, CNPT, AH2, Lec, Kir, Kiritsis, Ber, Tin, Jackiw, Tin1, RG, HHS, RG1, Gruzinov, HR, Cham4, Cham5, MC, BH, Muk1, Muk2, DW1, Gal1, ZWY, Muk3, Gal2, DW2},
and references therein.}

{}Within the perturbative framework, unitarity requires that, in order not to propagate a negative norm
state at the quadratic level in the action, the graviton mass
term be of the Fierz-Pauli form \cite{FP}. Furthermore, higher order terms should be such that they do not introduce
additional degrees of freedom that would destabilize the background \cite{BD}.\footnote{For recent related works, see, {\em e.g.}, \cite{AGS, AH1, CNPT, AH2, RG, RG1, HR, Cham4, Gal1, Gal2}, and references therein.}

{}In \cite{Unitarity}, following \cite{dS}, it was argued that perturbation theory appears to be inadequate, among other things, for the purposes of
addressing the issue of unitarity.\footnote{In \cite{Gal2} it is argued that in the massive gravity models of \cite{RG, RG1} perturbation theory breaks down already at a very low scale.} In \cite{Unitarity} a non-perturbative Hamiltonian analysis for the relevant metric
degrees of freedom was performed and it was argued that the full nonlinear theory is free of ghosts. The main result of \cite{Unitarity} is
that, in the gravitational Higgs mechanism in the Minkowski
background, {\em non-perturbatively} the Hamiltonian is bounded from below.\footnote{The same conclusion was reached in \cite{dS} in the de Sitter case. For prior works on massive gravity in the de Sitter space, see, {\em e.g.}, \cite{DN, Higuchi, DW, GI, GIS}, and references therein.} In fact, the results of \cite{Unitarity} are in complete agreement with the {\em full} Hamiltonian analysis performed in \cite{JK} for the original model proposed by 't Hooft \cite{thooft} (see below), in which the {\em full} (gauge-fixed) Hamiltonian is explicitly positive-definite and coincides with the Hamiltonian of \cite{Unitarity} for the relevant metric degrees of freedom.\footnote{In \cite{JK} the full Hamiltonian analysis also was performed for another model with non-polynomial ``potential" -- see subsection \ref{sq.rt.subsection} hereof for details. The full Hamiltonian analysis is technically challenging in general case. However, the Hamiltonian analysis of \cite{Unitarity} for the relevant metric degrees of freedom is tractable in other cases with nonlinear potentials, {\em e.g.}, the quadratic potential -- see subsection \ref{quad.subsection} hereof for details.}

{}In fact, this holds irrespective of whether the perturbative mass term is of the Fierz-Pauli form, including in the simplest case with
no higher-derivative couplings in the scalar sector,
first discussed in \cite{thooft}. Even though perturbatively the trace of the metric fluctuations is a propagating ghost, non-perturbatively the theory ``resums'' and the non-perturbative Hamiltonian is bounded from below.\footnote{However, neither \cite{Unitarity} nor this paper attempt to address the question of whether there is any superluminal propagation of signals or the related issue of (a)causality. In this regard, we emphasize
that the results of \cite{Gruzinov, DW1, DW2} are {\em intrinsically
perturbative} and do not appear to apply to the full
non-perturbative definition of the gravitational Higgs mechanism. To see if there is any superluminal propagation of signals in the full non-perturbative theory, it appears that one might have to
develop some new non-perturbative methods, which is clearly beyond the scope of this paper (and off the cuff it is not even evident what such non-perturbative methods would entail). The
non-Fierz-Pauli model of \cite{thooft} is the ``least non-perturbative" and might provide a fruitful testing
ground in this context. We also note that the issue of superluminal propagation does not appear to be relevant in the context of the application of the gravitational Higgs mechanism to a string theory description of QCD -- which was one of the primary motivations in 't Hooft's paper \cite{thooft} -- if QCD is to be described by string theory, all known consistent versions of which contain massless
gravity, then the graviton should presumably somehow acquire mass, and the gravitational Higgs mechanism is one way of approaching this problem.}

{}In this paper we further verify the results of \cite{Unitarity} by constructing exact non-perturbative massive solutions in the gravitational Higgs mechanism. These non-perturbative massive solutions do not appear to exhibit instabilities and describe vacuum configurations which are periodic in time. For example, in $D=3$ in the model of \cite{thooft}, we have purely longitudinal solutions with isotropic periodically expanding and contracting spatial dimensions with the metric ($0\leq \rho < 1$)
\begin{equation}\label{iso}
 ds^2 = -d\tau^2 + \left[1 - \rho~\cos(M\tau)\right]~\delta_{ij}dx^i dx^j~,
\end{equation}
where $M$ is the naive perturbative graviton mass.\footnote{Perturbatively, the traceless components of the graviton are positive-definite with mass $M$, while the trace component is a ghost with the same mass. Non-perturbatively, there is no ghost.} This is an exact solution. When the amplitude of the oscillations $\rho$ is large, the perturbative expansion breaks down as is evident upon examining the full nonlinear Hamiltonian ${\cal H}$ for the longitudinal mode:
\begin{equation}\label{H-intro}
 {{\cal H} \over 2\kappa M} = {M^2\over 2}~\sqrt{{2e^{-2q} - e^{-4q}\over{M^2 + 4(\partial_t q)^2}}} = {M\over 2} - {{(\partial_t q)^2 + M^2 q^2}\over M} +\dots~,
\end{equation}
which non-perturbatively is positive-definite (the first equality in (\ref{H-intro})), but perturbatively, when expanded in the weak-field approximation to the second
order in $q$ and $\partial_t q$, has a {\em fake} ghost of mass $M$ (the second equality in (\ref{H-intro})).\footnote{In (\ref{H-intro}) $q$ is a canonical variable for the longitudinal mode, $\kappa$ is a constant, $t$ is the time coordinate (related to but not the same as $\tau$ in (\ref{iso})) w.r.t. which the conjugate momentum for $q$ is defined, and the ellipses stand for higher order terms in $q$ and $\partial_t q$ (see $\S\S$ 4.2 and 6.1 for details).}

{}Also in the same model of \cite{thooft}, in any dimension $D$, we find exact solutions we refer to as ``cosmological strings", of the form:
\begin{equation}
 ds^2 = e^{\omega(t)}~\left[-dt^2 + (dx^1)^2\right] + \sum_{i=2}^{D-1} (dx^i)^2~.
\end{equation}
These solutions are 2-dimensional cosmological defects in a $D$-dimensional space-time with only one periodically expanding and contracting spatial dimension. The periodic function $\omega(t)$ has a period $T$, which depends on the oscillation amplitude. For small amplitudes $T\approx 2\pi/M$, where $M$ is the naive perturbative graviton mass.\footnote{The explicit form of $\omega$ is given in Subsections 7.1 and 7.2.}

{}Once higher-derivative terms are added in the scalar sector,\footnote{As was first discussed in \cite{ZK1}, higher-derivative terms are required to obtain the Fierz-Pauli term in the perturbative expansion.} there are other exact solutions, including, {\em e.g.}, $D=3$ purely non-longitudinal (``traceless") periodically expanding and contracting solutions with constant spatial volume with the metric of the form
\begin{eqnarray}
 &&ds^2 = -d\tau^2 + \left[1 + \rho~\cos(M_1\tau)\right](dx^1)^2 + \left[1 - \rho~\cos(M_1\tau)\right](dx^2)^2 + \nonumber\\
 &&\,\,\,2\rho~\sin(M_1\tau)~(dx^1)(dx^2)~,
\end{eqnarray}
where the amplitude of oscillations $\rho$ and the mass parameter $M_1$ are determined by the higher-derivative couplings in the scalar sector (see Subsection 7.3 for details).

{}Upon gauging away the scalar degrees of freedom, the simplest non-perturbative massive gravity action (in the Minkowski background) can be written as
\begin{equation}
 S_{MG} = M_P^{D-2}\int d^Dx \sqrt{-G}\left[ R + \mu^2\left(D - 2 - G^{MN}\eta_{MN}\right)\right]~,
\end{equation}
which corresponds to the model of \cite{thooft} without higher-derivative couplings in the scalar sector.\footnote{See Section 2 for the discussion of more general actions that allow for the Fierz-Pauli mass term in the linearized gravity.} This action, albeit nonlinear, allows to obtain various exact solutions, as we have done in this paper.\footnote{It would also be interesting to construct black hole solutions in this model.} Since non-perturbatively there is no ghost \cite{Unitarity}, it might be possible to utilize this model in pursuing one of the original motivations of \cite{thooft}, namely, string theory description of QCD, where massless spin-2 modes somehow must acquire mass.\footnote{Because of the tachyon problem, it is difficult to speculate about embedding the gravitational Higgs mechanism into bosonic string theory. However, such embedding appears to be both straightforward and natural in the context of supersymmetric string theory, where typically there is an abundance of scalars with flat directions (and no non-derivative couplings). (In this context, the role of the time-like scalar is played by a pseudoscalar dual to a $p$-form with $p=D-2$.) In supersymmetric string theory there appears to exist no evident obstruction to the aforesaid scalars acquiring coordinate-dependent vacuum expectation values thereby spontaneously breaking the diffeomorphism invariance and resulting in the gravitational Higgs mechanism. Such a string background would, however, appear to be non-perturbative (strongly coupled, and, in fact, time-dependent), which bodes well with the apparent non-perturbative nature of the gravitational Higgs mechanism. In this regard, let us note a difference between the gravitational Higgs mechanism and its gauge theory counterpart. In the latter scalar vacuum expectation values are constant, while in
the former they depend linearly on space-time coordinates. In fact, the background is not
even static. It would take infinite energy to destabilize such a background, which should therefore be stable. This is reminiscent to
infinite-tension domain walls discussed in \cite{DS,ZK3}.} Another motivation for massive gravity -- its large-scale modification -- is cosmology, including in the context of the currently observed accelerated expansion of the Universe \cite{Nova,Nova1} and the cosmological constant.

{}As an aside we also discuss massive solutions in New Massive Gravity \cite{BHT}. While such solutions are present in the linearized theory, we argue that already at the next-to-linear (quadratic) order in the equations of motion (and, more generally, for weak-field configurations) there are no massive solutions.

{}The rest of the paper is organized as follows. In Sections 2 and 3 we discuss
the gravitational Higgs mechanism in a general background, which results in linearized
massive gravity with the Fierz-Pauli mass term for the appropriately tuned
cosmological constant. In Section 4 we discuss non-perturbative massive gravity via the gravitational Higgs mechanism and review the non-perturbative Hamiltonian for
the relevant metric modes and its positive-definiteness. In Section 5 we discuss restrictions on higher curvature terms that we use in our discussion of New Massive Gravity in Section 8. In Section 6 we discuss non-perturbative massive solutions corresponding to the longitudinal mode. In Section 7 we discuss non-perturbative non-longitudinal massive solutions. In Section 8 we discuss massive solutions in New Massive gravity.

\section{Gravitational Higgs Mechanism}

{}The goal of this section is to obtain massive gravity in a general background via the gravitational Higgs mechanism. This is a generalization of the setup of \cite{dS} for the gravitational Higgs mechanism in the de Sitter space to a general background. Thus, let
\begin{equation}\label{action.0}
 S = S_G + {1\over 2} \int d^Dx ~G_{MN} T^{MN} ~,
\end{equation}
where
\begin{equation}
 S_G \equiv M_P^{D-2}\int d^Dx \sqrt{-G}\left[ R - {\widetilde \Lambda} + {\cal O}\left(R^2\right)\right]
\end{equation}
is an arbitrary generally covariant action constructed from the metric $G_{MN}$ and its derivatives, ${\widetilde \Lambda}$ is the cosmological constant, ${\cal O}\left(R^2\right)$ stands for higher curvature terms constructed from $R$, $R_{MN}$ and $R_{MNST}$, and we have included a coupling to the conserved energy momentum tensor $T^{MN}$:
\begin{equation}
 \nabla_N T^{NM} = 0~.
\end{equation}
Let ${\widetilde G}_{MN}$ be a background solution to the equations of motion corresponding to (\ref{action.0}):
\begin{equation}
 R_{MN} - {1\over 2} G_{MN} \left[R - {\widetilde \Lambda}\right] + {\cal O}\left(R^2\right) = {M_P^{2-D}\over 2\sqrt{-G}} T_{MN}~,
\end{equation}
The background metric ${\widetilde G}_{MN}$ generally is a function of the coordinates $x^S$: ${\widetilde G}_{MN} = {\widetilde G}_{MN}(x^0,\dots, x^{D - 1})$.

{}Let us now introduce $D$ scalars $\phi^A$. Let us normalize them such that they have dimension of length. Let us define a metric $Z_{AB}$ for the scalars as follows:
\begin{equation}
 Z_{AB}(\phi^0,\dots,\phi^{D-1}) \equiv {\delta_A}^M {\delta_B}^N {\widetilde G}_{MN}(\phi^0,\dots, \phi^{D - 1})~,
\end{equation}
{\em i.e.}, we substitute the space-time indices $M$ and $N$ in ${\widetilde G}_{MN}$ with the global scalar indices $A$ and $B$, and substitute $x^0\rightarrow \phi^0,\dots ,x^{D-1}\rightarrow \phi^{D-1}$ in the functional form of ${\widetilde G}_{MN}$.

{}Next, consider the induced metric for the scalar sector:
\begin{equation}
 Y_{MN} = Z_{AB} \nabla_M\phi^A \nabla_N\phi^B~.
\end{equation}
Also, let
\begin{equation}\label{Y}
 Y\equiv Y_{MN}G^{MN}~.
\end{equation}
The following action, albeit not most general,\footnote{\label{foot}One can consider a
more general setup with the scalar action constructed not just from $Y$,
but from $Y_{MN}$, $G_{MN}$ and $\epsilon_{M_0\dots M_{D-1}}$, see, {\em e.g.},
\cite{ZK1,Demir,Cham1,Oda2}. However, a simple action containing a scalar
function $V(Y)$ suffices to capture all qualitative features of the gravitational
Higgs mechanism.} will suffice for our purposes here:
\begin{equation}
 S_Y = M_P^{D-2}\int d^Dx \sqrt{-G}\left[ R - \mu^2 V(Y) + {\cal O}\left(R^2\right)\right] + {1\over 2} \int d^Dx ~G^{MN} T_{MN} ~,
 \label{actionphiY}
\end{equation}
where {\em a priori} the ``potential" $V(Y)$ is a generic function of $Y$, and $\mu$ is a mass parameter, while the cosmological constant ${\widetilde \Lambda}$ is subsumed in the definition of $V(Y)$ (see below).

{}The equations of motion read:
\begin{eqnarray}
 \label{phiY}
 && \nabla^M\left(V^\prime(Y)Z_{AB} \nabla_M \phi^B\right) = {1\over 2} {\partial Z_{BC}\over \partial\phi^A} G^{MN} \nabla_M \phi^B~\nabla_N\phi^C ~V^\prime(Y)~,\\
 \label{einsteinY}
 && R_{MN} - {1\over 2}G_{MN} R + {\cal O}\left(R^2\right) - \mu^2\left[ V^\prime(Y) Y_{MN}
 -{1\over 2}G_{MN} V(Y)\right] = \nonumber\\
 && = {M_P^{2-D}\over 2\sqrt{-G}} T_{MN}~,
\end{eqnarray}
where prime denotes a derivative w.r.t. $Y$. Multiplying (\ref{phiY}) by $\nabla_S\phi^A$ and contracting indices, we
can rewrite the scalar equations of motion as follows:
\begin{equation}\label{phiY.1}
 \partial_M\left[\sqrt{-G} V^\prime(Y) G^{MN}Y_{NS}\right] - {1\over 2}\sqrt{-G} V^\prime(Y) G^{MN}\partial_S Y_{MN} = 0~.
\end{equation}
Since the theory possesses full diffeomorphism symmetry, (\ref{phiY.1}) and (\ref{einsteinY}) are not all independent but linearly related
due to Bianchi identities. Thus, multiplying (\ref{einsteinY}) by $\sqrt{-G}$, differentiating w.r.t. $\nabla^N$ and contracting indices we arrive
at (\ref{phiY.1}).

{}We are interested in finding solutions of the form:
\begin{eqnarray}\label{solphiY}
 &&\phi^A = {\delta^A}_M~x^M~,\\
 \label{solGY}
 &&G_{MN} = {\widetilde G}_{MN}~.
\end{eqnarray}
Since on this solution $Y_{MN} = {\widetilde G}_{MN}$ and $Y = D$, (\ref{phiY.1}) is automatically satisfied. Furthermore, (\ref{einsteinY}) implies that
\begin{equation}
 R_{MN} - {1\over 2} G_{MN} \left[R - {\widetilde \Lambda}\right] + {\cal O}\left(R^2\right) = {M_P^{2-D}\over 2\sqrt{-G}} T_{MN}~,
\end{equation}
provided that
\begin{equation}\label{cosm.const}
 V(D) - 2 V^\prime(D) = {\widetilde \Lambda} / \mu^2~.
\end{equation}
For {\em non-generic} potentials this condition can be constraining. For example, if $V(Y) = a + Y$ and ${\widetilde \Lambda} = 0$, we have $a = -(D-2)$ implying that the vacuum energy density $\Lambda \equiv \mu^2 V(0) = a \mu^2$ in the unbroken phase ($\phi^A \equiv 0$) must be negative. However, since $\mu$ is an arbitrary mass scale, for {\em generic} potentials $V(Y)$ the above condition is not constraining. In particular, generically there is no restriction on the vacuum energy density $\Lambda = \mu^2 V(0)$ in the unbroken phase, which can be positive, negative or zero, even if one requires that the mass term for the graviton in the linearized theory is of the Fierz-Pauli form (see below).

\section{Linearized Massive Gravity}

{}In this section we discuss linearized fluctuations in the background given by (\ref{solphiY}) and (\ref{solGY}). Since diffeomorphisms are broken spontaneously, the equations of motion are invariant under the full diffeomorphism invariance. The scalar fluctuations $\varphi^A$ can therefore be gauged away using the diffeomorphisms:
\begin{equation}\label{diffphiY}
 \delta\varphi^A =\nabla_M \phi^A \xi^M = {\delta^A}_M ~\xi^M~.
\end{equation}
However, once we gauge away the scalars, diffeomorphisms
can no longer be used to gauge away any of the graviton components $h_{MN}$ defined as:
\begin{equation}
 G_{MN} = {\widetilde G}_{MN} + h_{MN}~,
\end{equation}
where ${\widetilde G}_{MN}$ is the background metric defined in the previous section. We will use the notation $h \equiv {\widetilde G}^{MN} h_{MN}$.

{}After setting $\varphi^A = 0$, we have
\begin{eqnarray}
 && Y_{MN} = {\widetilde G}_{MN}~,\\
 && Y = Y_{MN} G^{MN} = D - h + {\cal O}\left(h^2\right) ~,
\end{eqnarray}
Due to diffeomorphism invariance, the scalar equations of motion (\ref{phiY}) are related to (\ref{einsteinY}) via Bianchi identities. We will therefore focus on (\ref{einsteinY}). Linearizing the terms containing $V$, we obtain:
\begin{eqnarray}
 &&R_{MN} - {1\over 2}G_{MN} \left[R - {\widetilde \Lambda}\right] + {\cal O}\left(R^2\right)- \nonumber\\
 && -{M^2\over 2} \left[{\widetilde G}_{MN} h - \zeta h_{MN}\right] + {\cal O}\left(h^2\right) = \nonumber\\
 && = {M_P^{2-D}\over 2\sqrt{-G}} T_{MN}~,\label{lin.eom}
\end{eqnarray}
where
\begin{eqnarray}
 &&M^2 \equiv \mu^2 \left[V^\prime(D) - 2 V^{\prime\prime}(D)\right]~,\label{M}\\
 &&\zeta M^2 \equiv 2\mu^2 V^\prime(D)~.\label{zeta}
\end{eqnarray}
This corresponds to adding a graviton mass term of the form
\begin{equation}
 -{M^2\over 4} \left[\zeta h_{MN}h^{MN} - h^2\right]
\end{equation}
to the gravity action $S_G$,
and the Fierz-Pauli combination corresponds to taking $\zeta = 1$. This occurs for a special class of potentials with
\begin{equation}\label{tune-V}
 V^\prime(D) = -2V^{\prime\prime}(D)~.
\end{equation}
Thus, as we see, we can obtain the Fierz-Pauli combination of the mass term
for the graviton if we tune {\em one} combination of couplings. In fact, this tuning is nothing but the tuning of
the vacuum energy density $\Lambda = \mu^2 V(0)$ in the unbroken phase -- indeed, (\ref{tune-V}) relates the vacuum energy density in the unbroken phase to higher derivative couplings.

\subsection{Linear Potential}

{}The simplest potential is given by \cite{thooft}:
\begin{equation}
 V = a + Y
\end{equation}
From (\ref{cosm.const}) we have
\begin{equation}\label{a.lin}
 a = {{\widetilde \Lambda}\over \mu^2} - (D - 2)
\end{equation}
The vacuum energy density in the unbroken phase $\Lambda = \mu^2 V(0)$ is negative for Minkowski solutions. Also, for this potential (\ref{tune-V}) cannot be satisfied, {\em i.e.}, we cannot have the Fierz-Pauli mass term, which requires $\zeta = 1$ in (\ref{zeta}) and instead we have $\zeta = 2$. Perturbatively, the trace $h \equiv {\widetilde G}^{MN} h_{MN}$ is a propagating ghost degree of freedom. However, as was argued in \cite{Unitarity}, non-perturbatively there is no ghost and the Hamiltonian for the relevant degrees of freedom (see below) is bounded from below. To achieve the Fierz-Pauli mass term, higher derivative terms for scalars are needed \cite{ZK1}.

\subsection{Quadratic Potential}

{}Thus, consider a simple example:
\begin{equation}\label{quad.pot}
 V = a + Y + \lambda Y^2~.
\end{equation}
The first term corresponds to the vacuum energy density $\Lambda = \mu^2 V(0)$ in the unbroken phase, the second term is the kinetic term for the scalars (which can always be normalized such that the corresponding coefficient is 1 by adjusting $\mu$), and the third term is a four-derivative term. From (\ref{tune-V}) we then have:
\begin{equation}
 \lambda = -{1\over{2(D+2)}}~,
\end{equation}
and the graviton mass is given by:
\begin{equation}
 M^2 = {4\mu^2\over{D+2}}~.
\end{equation}
Moreover, from (\ref{cosm.const}) we have:
\begin{equation}
 a = {{\widetilde \Lambda}\over \mu^2} - {{D^2 + 4D - 8}\over {2(D+2)}}~,
\end{equation}
which is nothing but tuning of the vacuum energy density $\Lambda = \mu^2 V(0)$ in the unbroken phase against the higher derivative coupling $\mu^2 \lambda$.

{}Here the following remark is in order. In the above example with the quadratic potential (\ref{quad.pot}) the vacuum energy density $\Lambda = \mu^2 V(0)$ in the unbroken phase must be negative in the case of the Minkowski background (${\widetilde \Lambda} = 0$). However, generically, there is no restriction on $\Lambda$, even in the case of the Minkowski background, if we allow cubic and/or higher order terms in $V(Y)$, or consider non-polynomial $Y(Y)$.

\subsection{Exponential Potential}

{}Thus, consider another simple example:
\begin{equation}\label{exp.pot}
 V = a + b~e^{-\lambda Y}~.
\end{equation}
Then from (\ref{tune-V}) we then have:
\begin{equation}\label{FP.exp}
 \lambda = {1\over 2}~,
\end{equation}
and from (\ref{cosm.const}) we have:
\begin{equation}
 a = {{\widetilde \Lambda}\over \mu^2} - 2\, b\, e^{-D/2}~.
\end{equation}
In this case the vacuum energy density in the unbroken phase $\Lambda = \mu^2 V(0) = a + b$ can therefore be positive, negative or zero depending on the value of the parameter $b$ irrespective of the value of ${\widetilde \Lambda}$.

\subsection{Square-root Potential}

{}Finally, consider the following non-polynomial potential:
\begin{equation}\label{sq.rt}
 V = a + \sqrt{Y + b}~.
\end{equation}
From (\ref{tune-V}) we have:
\begin{equation}
 b = -(D-1)~,
\end{equation}
and from (\ref{cosm.const}) we further have
\begin{equation}
 a = {{\widetilde \Lambda}\over \mu^2}~.
\end{equation}
An interesting feature of this square-root potential is that there is no unbroken phase as we must have $Y\geq (D-1)$.

\section{Non-perturbative Massive Gravity}

{}The gravitational Higgs mechanism provides a non-perturbative definition of massive gravity in a general background. Thus, if we use diffeomorphisms to gauge away the scalars by fixing them to their background values (\ref{solphiY}), the full non-perturbative action is then given by
\begin{equation}
 S_{MG} = M_P^{D-2}\int d^Dx \sqrt{-G}\left[ R - \mu^2 V(G^{MN}{\widetilde G}_{MN}) + {\cal O}\left(R^2\right)\right] + {1\over 2} \int d^Dx ~G^{MN} T_{MN} ~,
 \label{massivegravity}
\end{equation}
which describes massive gravity in the background metric ${\widetilde G}_{MN}$. The equation of motion reads
\begin{eqnarray}\label{R.eom}
 && R_{MN} - {1\over 2}G_{MN} R + {\cal O}\left(R^2\right) - \mu^2\left[ V^\prime(G^{ST}{\widetilde G}_{ST}) {\widetilde G}_{MN}
 -{1\over 2}G_{MN} V(G^{ST}{\widetilde G}_{ST})\right] = \nonumber\\
 && = {M_P^{2-D}\over 2\sqrt{-G}} T_{MN}~,
\end{eqnarray}
with the Bianchi identity
\begin{equation}\label{phi.eom}
 \partial_M\left[\sqrt{-G} V^\prime(G^{RT}{\widetilde G}_{RT}) G^{MN}{\widetilde G}_{NS}\right] - {1\over 2}\sqrt{-G} V^\prime(G^{RT}{\widetilde G}_{RT}) G^{MN}\partial_S {\widetilde G}_{MN} = 0~,
\end{equation}
which is equivalent to the gauge-fixed equations of motion for the scalars.

\subsection{An Example: Schwarzschild Background}

{}This construction allows to define massive gravity in nontrivial curved backgrounds. The de Sitter background was discussed in \cite{dS}. Another example is the Schwarzschild background, for which the background metric ${\widetilde G}_{MN}$ is given by:
\begin{equation}
 ds^2 = {\widetilde G}_{MN} dx^A dx^B = -A^2 dt^2 + B^2 dr^2 + C^2 \gamma_{ab} dx^a dx^b~,
\end{equation}
where
\begin{eqnarray}
 &&A = B^{-1} = \sqrt{1 - {r_* / r}}~,\\
 &&C = r~,
\end{eqnarray}
and $\gamma_{ab}$ is a metric on the unit sphere $S^{d-1}$, $d = D - 1$. This, provides an explicit construction of massive gravity in the Schwarzschild background (not to be confused with a construction of Schwarzschild-like solutions in massive gravity).

\subsection{Unitarity: Minkowski Background}

{}Following \cite{dS} and \cite{Unitarity}, in this subsection we discuss unitarity of massive gravity via the gravitational Higgs mechanism in the Minkowski background by studying the full non-perturbative Hamiltonian for conformal and helicity-0 modes. This suffices to deduce whether there are any additional degrees of freedom that destabilize the background, assuming that transverse-traceless graviton components are positive-definite.\footnote{In some cases with higher curvature terms transverse-traceless graviton components are not necessarily positive-definite. In such cases our discussion here assumes that no higher-curvature terms are present.}

{}To identify the relevant modes in the full nonlinear theory, let us note that
in the linearized theory the potentially ``troublesome'' mode is the
longitudinal helicity-0 mode $\rho$. However, we must also include the
conformal mode $\omega$ as there is kinetic mixing between $\rho$ and $\omega$.
In fact, $\rho$ and $\omega$ are not independent but are related via Bianchi
identities. Therefore, in the linearized language one must look at the modes of
the form\footnote{However, as was discussed in detail in \cite{dS,Unitarity}, this naive intrinsically perturbative parametrization is inadequate, among other things, for addressing the issue of unitarity.}
\begin{equation}\label{param.lin}
 h_{MN} = \eta_{MN}~\omega + \nabla_M\nabla_N \rho~.
\end{equation}
Furthermore, based on symmetry considerations, namely, the $SO(D-1)$ invariance
in the spatial directions,\footnote{Indeed, negative norm states cannot arise from purely space-like components or spatial derivatives, and are due to time-like components and/or time derivatives.} we can focus on field configurations independent of
spatial coordinates \cite{dS, Unitarity}. Indeed, for our purposes here we can compactify the
spatial coordinates on a torus $T^{D-1}$ and disregard the Kaluza-Klein modes.
This way we reduce the $D$-dimensional theory to a classical mechanical system,
which suffices for our purposes here. Indeed, with proper care (see \cite{dS}), if
there is a negative norm state in the uncompactified theory, it will be visible
in its compactified version, and vice-versa.

{}Let us therefore consider field configurations of the form:
\begin{equation}\label{param.full}
 G^{MN} = {\rm diag}(g(t)~\eta^{00}, f(t)~\eta^{ii})~,
\end{equation}
where $g(t)$ and $f(t)$ are functions of time $t$ only. The action
(\ref{massivegravity}) then reduces as follows:\footnote{Here we omit the source term as it does not affect the unitarity analysis.}
\begin{eqnarray}\label{compact}
 S_{MG} = -\kappa \int dt ~g^{-{1\over 2}} f^{-{{D-1}\over 2}}
\left\{Q(gU^2) + \mu^2~V(g + \Omega)\right\}~,
\end{eqnarray}
where
\begin{eqnarray}
 &&\kappa\equiv {M_P^{D-2} W_{D-1}}~,\\
 &&Q(x)\equiv \sum_{k=1}^{\infty} c_k x^k~,\label{sumk}\\
 &&c_1 \equiv (D-1)(D-2)~,\\
 &&U\equiv {1\over 2}\partial_t\ln(f)~,\\
 &&\Omega\equiv (D-1)f~,
\end{eqnarray}
and $W_{D-1}$ is the volume in the spatial dimensions ({\em i.e.}, the volume
of $T^{D-1}$). Also, the $k=1$ term in (\ref{sumk}) corresponds to the Einstein-Hilbert term, while the $k>1$ terms (the sum over $k$, {\em a priori}, can be finite or infinite) correspond to the ${\cal O}\left(R^2\right)$ terms and we are assuming that the ${\cal O}\left(R^2\right)$ are such that only first time-derivatives of $f$ appear, which imposes certain conditions on the higher curvature terms (see below). Also, note that $g$ is a Lagrange multiplier. We can integrate
out $g$ and obtain the corresponding action for $f$. It is then this action that we need to test for the presence of a negative norm state.

{}However, there is a simple way to see if the Hamiltonian is positive-definite. The equation of motion for $g$ reads:
\begin{equation}\label{geq}
 \mu^2\left[V(g+\Omega) - 2gV^\prime(g +\Omega)\right] = 2gU^2 Q^\prime(gU^2) - Q(gU^2)~.
\end{equation}
For our purposes here it is more convenient to work with the canonical
variable $q$, where
\begin{eqnarray}
 &&q\equiv {1\over 2} \ln(f)~,\\
 &&\Omega = (D-1)e^{2q}~,\\
 &&U = \partial_t q~,
\end{eqnarray}
and the action reads:
\begin{eqnarray}\label{compact1}
 S_{MG} = \int dt~L = -\kappa \int {dt} ~g^{-{1\over 2}} e^{-(D-1)q}
\left\{Q(g U^2) + \mu^2~V(g + \Omega)\right\}~,
\end{eqnarray}
where $L$ is the Lagrangian. This action corresponds to a classical mechanical
system with a lagrange multiplier $g$.

{}Next, the conjugate momentum is given by
\begin{equation}
 p = {{\partial L} \over {\partial(\partial_t q)}} = -2\kappa e^{-(D-1)q} g^{{1\over 2}} U Q^\prime(gU^2)~,
\label{momentum}
\end{equation}
where we have used (\ref{geq}) to eliminate terms containing
\begin{equation}
{\hat g} \equiv {{\partial g} \over {\partial(\partial_\tau q)}}~,
\end{equation}
and the Hamiltonian is given by
\begin{equation}\label{Hamiltonian}
 {\cal H} = p~\partial_t q - L = 2 \kappa\mu^2 g^{{1\over 2}} e^{-(D-1)q}V^\prime(g + \Omega)~,
\end{equation}
where we again used (\ref{geq}).

{}Actually, this Hamiltonian can be obtained up to a normalization constant from (\ref{phiY.1}), which for the field configurations (\ref{param.full}) reduces to
\begin{equation}
 \partial_t\left[g^{{1\over 2}} e^{-(D-1)q}V^\prime(g + \Omega)\right]=0~.
\end{equation}
This is nothing but the condition that the Hamiltonian is constant.

\subsubsection{Linear Potential}

{}The Hamiltonian is evidently positive-definite for the linear potential $V = a + Y$ of \cite{thooft},\footnote{In \cite{thooft} it was observed that perturbatively the theory was non-unitary.} in complete agreement with the full Hamiltonian analysis of \cite{JK} (see \cite{Unitarity} for details). As was pointed out in \cite{Unitarity}, the absence of ghosts in the full non-perturbative theory despite the fact that for the linear potential we do not have the Fierz-Pauli mass-term, is yet another illustrative example of pitfalls of linearization.

\subsubsection{Quadratic Potential}\label{quad.subsection}

{}The proof that the Hamiltonian is positive-definite for the quadratic potential (\ref{quad.pot}) is given in \cite{Unitarity} for the Einstein-Hilbert case ($c_{k>1} = 0$ in (\ref{sumk})). The proof in the case of general $Q(gU^2)$ is essentially the same with the substitution $c_1 gU^2 \rightarrow Q(gU^2)$ assuming that $c_{k>1} \geq 0$.

\subsubsection{Exponential Potential}

{}The Hamiltonian is evidently positive-definite for the exponential potential (\ref{exp.pot}) with $b\lambda < 0$. In fact, ghosts are absent even without requiring (\ref{FP.exp}), which is the requirement that the linearized mass term be of the Fierz-Pauli form.

\subsubsection{Square-root Potential}\label{sq.rt.subsection}

{}The Hamiltonian is evidently positive-definite for the square-root potential (\ref{sq.rt}). This is consistent with the full Hamiltonian analysis of \cite{JK} for this potential.\footnote{The Hamiltonian analysis of \cite{JK} for this potential apparently assumes $b>0$. However, it appears that it can also be performed for $b<0$.}

\subsection{Unitarity: de Sitter Background}

{}The unitarity argument for the de Sitter background closely parallels that for the Minkowski background. In the de Sitter case we consider field configurations of the form:
\begin{equation}\label{param.full.dS}
 G^{MN} = {\rm diag}(g(t)~{\widetilde G}^{00}, f(t)~{\widetilde G}^{ii})~,
\end{equation}
where $g(t)$ and $f(t)$ are functions of time $t$ only, and ${\widetilde G}_{MN}$ is the de Sitter metric, which we can choose as follows:
\begin{equation}
 {\widetilde G}_{MN} = {\rm diag}(\eta_{00}, e^{-2Ht}\eta_{ii})~,
\end{equation}
where $H$ is the Hubble parameter. The above unitarity argument for the Minkowski background is essentially unchanged if we define
\begin{equation}
 q\equiv {1\over 2}\ln(f) + Ht
\end{equation}
as the canonical variable.

\section{Restrictions on Higher Curvature Terms}

{}In the previous section we assumed that (\ref{sumk}) does not contain higher derivative terms. This imposes restrictions on higher curvature terms. Here we discuss these restrictions in the case of terms quadratic in curvature. The most general quadratic curvature contribution to the action is given by:
\begin{equation}
M_P^{D-2}\int d^Dx \sqrt{-G}\left[\alpha R^2 + \beta R^{MN} R_{MN} + \gamma\left(R^{MNST}R_{MNST} - 4R^{MN}R_{MN} + R^2\right)\right]~.
\end{equation}
The Gauss-Bonnet combination does not introduce higher derivatives. The other two terms do not contribute higher derivative terms provided that
\begin{equation}\label{alphabeta}
 4(D-1)\alpha + D\beta = 0~,
\end{equation}
and in this case we have ($c_{k>2} = 0$):
\begin{equation}
 c_2 = {(D-1) (D-2) (D-4)\over 12}~\left[4(D-3)\gamma - (D-2)\beta\right]~.
\end{equation}
Note that $c_2$ vanishes for $D=4$. The same is the case for $D>4$ and
\begin{equation}
 \beta={4(D-3)\gamma\over(D-2)}~,
\end{equation}
when the quadratic curvature contribution is of the form:
\begin{equation}
M_P^{D-2}\int d^Dx \sqrt{-G} ~C^{MNST}C_{MNST}~,
\end{equation}
where
$C_{MNST}$ is the Weyl tensor and the theory is at a critical point discussed in \cite{crit} with a unique vacuum.

{}Here we should point out that our discussion of higher-curvature terms in this section is motivated primarily by our discussion of New Massive Gravity in Section 8 (as opposed to in the context of the gravitational Higgs mechanism).

\section{Non-perturbative Longitudinal Solutions}

{}In this section we discuss non-perturbative solutions to the equations of motion in the Minkowski background with the aim to further demonstrate that, despite apparent issues with perturbative unitarity, {\em e.g.}, in the case of the linear potential where perturbatively there is a propagating ghost, non-perturbatively no negative-norm states appear and the background appears to be stable. In most cases solving highly nonlinear equations of motion analytically is challenging. However, there are two cases of interest where the equations of motion are tractable. In this section we will assume that no higher-curvature terms are present.

\subsection{Linear Potential}

{}Let us start with the linear potential. From (\ref{geq}) we have
\begin{equation}
 g = {{\Omega - (D-2)}\over {1 + c_1~{U^2\over\mu^2}}}~,
\end{equation}
where we have used (\ref{a.lin}) together with ${\widetilde \Lambda} = 0$. Furthermore, from the condition that the Hamiltonian is constant, we have:
\begin{equation}\label{g.lin}
 g~e^{-2(D-1)q} = \nu^2~,
\end{equation}
where
\begin{equation}\label{nu}
 \nu \equiv {{\cal H}\over{2\kappa\mu^2}} > 0~.
\end{equation}
We therefore have the following first-order equation for $q$:
\begin{equation}\label{q.lin}
 (D-1)(D-2)~{\nu^2\over\mu^2}~\left(\partial_t q\right)^2 = (1 - \nu^2) - (e^{-2q} - 1)^2~\sum_{k=0}^{D-3} (k+1)~e^{-2kq}~,
\end{equation}
where we have used the following identity
\begin{equation}
 n~x^{n-1} - (n-1)~x^n = 1 - (x-1)^2~\sum_{k=0}^{n-2}(k+1)~x^k
\end{equation}
along with the definitions of $\Omega$ and $c_1$. Note that from (\ref{q.lin}) we have $\nu\leq 1$.

{}The equation of motion (\ref{q.lin}) can be solved in the $D=3$ case, which suffices for our purposes here:
\begin{equation}
 e^{2q} = {1\over \nu^2}\left[1+\sqrt{1-\nu^2}~\cos\left(\sqrt{2}\mu(t-t_0)\right)\right]~,
\end{equation}
where $t_0$ is an integration constant. Note that, due to (\ref{g.lin}), we have
\begin{equation}
 g^{1\over 2} = \nu~e^{2q}~.
\end{equation}
The metric is given by
\begin{equation}
 ds^2 = -g^{-1}~dt^2 + e^{-2q}~\delta_{ij}dx^i dx^j~.
\end{equation}
Let us redefine the time coordinate as follows:
\begin{equation}\label{tau-t}
 d\tau \equiv g^{-{1\over 2}}~dt~.
\end{equation}
Then we have
\begin{equation}
 \tau - \tau_0 = {1\over{\sqrt{2}\mu}}~\arccos\left({{\sqrt{1-\nu^2} + \cos\left(\sqrt{2}\mu(t-t_0)\right)}\over{1 + \sqrt{1-\nu^2}~\cos\left(\sqrt{2}\mu(t-t_0)\right)}}\right)~,
\end{equation}
where $\tau_0$ is an integration constant. In terms of this new time coordinate $\tau$ we have:
\begin{equation}
 ds^2 = -d\tau^2 + e^{-2q}~\delta_{ij}dx^i dx^j~,
\end{equation}
where
\begin{equation}
 e^{-2q} = 1 - \sqrt{1-\nu^2}~\cos\left(\sqrt{2}\mu(\tau-\tau_0)\right)~.
\end{equation}
The non-perturbative Hamiltonian for the linear potential in general dimension $D$ is given by:
\begin{equation}
 {\cal H}_{\mbox{\small{non-pert}}} = 2\kappa\mu^3~\sqrt{{(D-1)e^{-2(D-2)q} - (D-2)e^{-2(D-1)q}\over{\mu^2 + (D-1)(D-2)(\partial_t q)^2}}}~.
\end{equation}
If we naively expand the Hamiltonian in the weak-field approximation to the second order in $q$ and $\partial_t q$, we obtain:
\begin{equation}
 {\cal H}_{\mbox{\small{pert}}} = 2\kappa\left\{\mu^2 - (D-1)(D-2)\left[{1\over 2}\left(\partial_t q\right)^2 + \mu^2~q^2\right] + \dots\right\}~,
\end{equation}
where the ellipses stand for higher order terms in $q$ and $\partial_t q$. This naive expansion produces a ghost of mass $M_* = \sqrt{2}\mu$, which is precisely the mass of the perturbative would-be ghost -- the trace $h$ -- in (\ref{lin.eom}) when $\zeta = 2$.\footnote{\label{foot2}For general $\zeta\not=1$, perturbatively the mass of the propagating trace component $h$ is given by $M_*^2 = \zeta(D-\zeta)\mu^2/(D-2)(\zeta - 1)$ -- see, {\em e.g.}, \cite{ZK1} for details.} However, non-perturbatively there is no ghost as the Hamiltonian is positive. Instead, we have non-perturbative oscillations with the same mass parameter $M_*$, but with the amplitude controlled by $\nu$. This amplitude is small for $\nu = 1 - \epsilon$, where $\epsilon\ll 1$. However, if $\nu$ is not close to 1, then the perturbative expansion breaks down. Therefore, the presence of a ghost in the perturbative expansion is merely an artifact of linearization.

{}For the sake of completeness, let us note that the equation of motion for general $D$ in terms of the time coordinate $\tau$ reads:
\begin{equation}
 (D-1)(D-2)~{e^{-2(D-1)q}\over\mu^2}~\left(\partial_\tau q\right)^2 = (1 - \nu^2) - (e^{-2q} - 1)^2~\sum_{k=0}^{D-3} (k+1)~e^{-2kq}~,
\end{equation}
with
\begin{equation}
 g^{1\over 2} = \nu~e^{(D-1)q}~,
\end{equation}
and $\tau$ is related to $t$ via (\ref{tau-t}).

\subsection{Square-root Potential}

{}As was argued in \cite{Unitarity}, in the case of the quadratic potential the Fierz-Pauli point is not special in any way as far as non-perturbative unitarity is concerned, in fact, non-perturbatively the Hamiltonian is bounded from below for a continuous range of values of the four-derivative coupling $\lambda$ in (\ref{quad.pot}) smoothly interpolating between the linear potential ($\zeta = 2$ in (\ref{lin.eom})) and the Fierz-Pauli point ($\zeta=1$ in (\ref{lin.eom})). The aim of this subsection is to understand if there is anything ``special" happening at the Fierz-Pauli point, if not from the non-perturbative unitarity standpoint, at least at the level of the non-perturbative equations of motion. The latter are cumbersome to analyze for the quadratic potential. However, they are tractable for the square-root potential (\ref{sq.rt}) in the Minkowski background at the Fierz-Pauli point, {\em i.e.}, $a=0$ and $b=-(D-1)$.

{}From (\ref{geq}) we have
\begin{equation}
  {(D-1)(f-1)\over\sqrt{g + (D-1)(f-1)}}= c_1~{gU^2\over\mu^2}~.
\end{equation}
Furthermore, from the condition that the Hamiltonian is constant, we have:
\begin{equation}\label{g.sq}
 {g^{1\over 2}~f^{-{{D-1}\over 2}} \over 2 \sqrt{g + (D-1)(f-1)}} = \nu~,
\end{equation}
where $\nu$ is defined in (\ref{nu}). Let us define a new time coordinate ${\widetilde t}$ via
\begin{equation}
 d{\widetilde t} = g^{-3/4} dt~,
\end{equation}
We then have the following equation for $f$:
\begin{equation}
 (f-1)f^{{D+3}\over 2} = {{D - 2}\over 8\nu\mu^2} ~\left(\partial_{\widetilde t} f\right)^2~.
\end{equation}
Let us define
\begin{equation}
 {\widetilde f} \equiv f^{-1}~.
\end{equation}
We have
\begin{equation}\label{tilde.f}
 (1-{\widetilde f}){\widetilde f}^{{3-D}\over 2} = {{D - 2}\over 8\nu\mu^2} ~\left(\partial_{\widetilde t} {\widetilde f}\right)^2~.
\end{equation}
One solution is ${\widetilde f} \equiv 1$, {\em i.e.}, $f\equiv 1$, which implies $\nu = 1/2$. Note that $g$ is arbitrary in this case and can be absorbed into the definition of the time coordinate.

{}To see if there are any other solutions, let us consider the $D=3$ case, which suffices for our purposes here. In $D=3$ a nontrivial solution to (\ref{tilde.f}) is given by
\begin{equation}\label{sol.tilde.f}
 {\widetilde f} = 1 - 2\nu\mu^2~\left({\widetilde t} - {\widetilde t}_0\right)^2~,
\end{equation}
where ${\widetilde t}_0$ is an integration constant. This implies that ${\widetilde f} \leq 1$ and $f \geq 1$. Note that the metric is given by
\begin{equation}
 ds^2 = -g^{1\over 2} d{\widetilde t}^2 + {\widetilde f}~\delta_{ij} dx^i dx^j~,
\end{equation}
and from (\ref{g.sq}) we have
\begin{equation}
 g = {{8\nu^2 f^2 (f - 1)}\over{1 - 4\nu^2 f^2}}~,
\end{equation}
so we must have $f < 1/(2\nu)$ and ${\widetilde f} > 2\nu$, {\em i.e.}, $\nu < 1/2$.

{}On the other hand, from (\ref{R.eom}) (with the higher curvature terms and the source term set to zero), we have for the scalar curvature
\begin{equation}
 R = \mu^2~{{D-1}\over {D-2}}~{{g + (D-1)f - D}\over\sqrt{g + (D-1)(f-1)}}~.
\end{equation}
So, in the above solution (\ref{sol.tilde.f}), when $f$ approaches $1/(2\nu)$ from below, which occurs at
\begin{equation}
 {\widetilde t}_{\pm} = {\widetilde t}_0 \pm {1\over \mu}~\sqrt{{1\over 2\nu} - 1}~,
\end{equation}
the scalar curvature $R$ diverges ($R\rightarrow +\infty$) as $g$ diverges and we have singularities at ${\widetilde t}_\pm$. Furthermore, as $f$ approaches 1 from above, which occurs when
\begin{equation}
 \left|{\widetilde t} -{\widetilde t}_0\right| \rightarrow 0~,
\end{equation}
the scalar curvature $R$ also diverges ($R\rightarrow -\infty$) as $g$ goes to zero.

{}As in the previous subsection, let us introduce the time coordinate $\tau$ defined via
\begin{equation}
 d\tau = g^{-{1\over 2}}dt = g^{1\over 4}d{\widetilde t}~.
\end{equation}
Then the metric reads
\begin{equation}
 ds^2 = -d\tau^2 + {\widetilde f}~\delta_{ij} dx^i dx^j~.
\end{equation}
Since
\begin{equation}
 g^{1\over 4} \sim \left|{\widetilde t} - {\widetilde t}_\pm\right|^{-{1\over 4}}
\end{equation}
as ${\widetilde t}$ approaches ${\widetilde t}_\pm$, it takes finite time $\tau$ to reach the singularities at ${\widetilde t} = {\widetilde t}_\pm$, which are therefore true (not coordinate) singularities. The singularity at ${\widetilde t} = {\widetilde t}_0$ is also a true singularity. Therefore, the solution (\ref{sol.tilde.f}) with $\nu < 1/2$ is a singular solution, which appears to be non-physical, and should be discarded.\footnote{Perhaps inclusion of higher-curvature terms could smooth out the singularities. Note that this solution passes through the epoch of vanishing (small) curvature.} This leaves us with the sole non-singular solution $f\equiv 1$ with $\nu = 1/2$.

{}Thus, what is ``special" at the Fierz-Pauli point is that non-perturbatively there appears to be no dynamics associated with the longitudinal mode. In contrast, as we saw in the previous subsection for the linear potential, at a non-Fierz-Pauli point we have oscillating non-perturbative solutions with the mass scale $M_* = \sqrt{2}\mu$, same as the mass of the would-be perturbative ghost; however, non-perturbatively there is no ghost and the Hamiltonian is bounded from below.

\subsection{Full Equations of Motion}

{}In this subsection we check that the solutions we found using the above Hamiltonian approach satisfy the full equations of motion. We are looking for the solutions of the form:
\begin{equation}\label{ua}
 G_{MN} = {\rm diag}(u(t)~\eta_{00}, e^{2A(t)}~\eta_{ii})~.
\end{equation}
Note the relation to our notations above: $u = g^{-1}$ and $A = -q$. We have:
\begin{eqnarray}
 &&R_{ij} = \left[{A^{\prime\prime}\over u} - {A^\prime u^\prime \over 2u^2} + {(D-1)\over u} (A^\prime)^2\right]~e^{2A}~\eta_{ij}~,\\
 &&R_{00} = -(D-1)\left[A^{\prime\prime} + (A^\prime)^2 - {A^\prime u^\prime \over 2u}\right]~,\\
 &&R_{i0} = 0~,
\end{eqnarray}
where a prime on $A$ and $u$ denotes a time derivative (not to be confused with a prime on the potential $V$ defied above). The equations of motion (\ref{R.eom}) and (\ref{phi.eom}) read (assuming the Minkowski background metric ${\widetilde G}_{MN} = \eta_{MN}$ and no source or higher curvature terms):
\begin{eqnarray}
 &&R_{MN} = \mu^2 \left[V^\prime(X)~\eta_{MN} +  {{V(X) - X~V^\prime(X)}\over {D-2}}~G_{MN}\right]~,\\
 &&\partial_M\left[\sqrt{-G}V^\prime(X) G^{MN}\right] = 0~,\label{Bianchi}
\end{eqnarray}
where
\begin{equation}
 X\equiv u^{-1} + (D-1)e^{-2A}~.
\end{equation}
Therefore, for the metric (\ref{ua}) we have:
\begin{eqnarray}
 &&A^{\prime\prime} - {A^\prime u^\prime \over 2u} + (D-1)(A^\prime)^2 = \mu^2 u \left[V^\prime(X)e^{-2A} + {{V(X) - X~V^\prime(X)}\over {D-2}}\right],\\
 &&(D-1)\left[A^{\prime\prime} + (A^\prime)^2 - {A^\prime u^\prime \over 2u}\right] = \mu^2 \left[V^\prime(X) + {{V(X) - X~V^\prime(X)}\over {D-2}}~u\right],\\
 &&u = {1\over \nu^2} e^{2(D-1)A}~\left[V^\prime(X)\right]^2~,
\end{eqnarray}
where $\nu$ is an integration constant and the last equation for $u$ follows from the Bianchi identity (\ref{Bianchi}). Combining these equations, we have:
\begin{eqnarray}
 &&(D-1)(D-2){\nu^2\over\mu^2}(A^\prime)^2 = \nonumber\\
 &&\,\,\, (D-1)V^\prime(X) e^{2(D-2)A} + \left[V(X) - X~V^\prime(X)\right] e^{2(D-1)A} - \nu^2 V^\prime(X)~.
\end{eqnarray}
For the linear potential $V(X) = X - (D-2)$ we recover (\ref{q.lin}). Furthermore, the second-order equation for $A$ reads
\begin{equation}
 A^{\prime\prime} = {\mu^2\over\nu^2}~\left[e^{2(D-2)A} - e^{2(D-1)A}\right]~,
\end{equation}
which follows from the first-order equation (\ref{q.lin}).

\section{Non-perturbative Non-longitudinal Solutions}

{}In the previous section we saw that non-perturbatively in the case of the linear potential (non-Fierz-Pauli point) we have oscillating solutions corresponding to the longitudinal mode with the mass parameter $M_* = \sqrt{2}\mu$, same as the mass of the would-be perturbative ghost; however, non-perturbatively there is no ghost, the Hamiltonian is bounded from below, and the aforementioned non-perturbative oscillating solutions appear to be well-behaved. In the case of the square-root potential (Fierz-Pauli point) there appears to be no dynamics associated with the longitudinal mode. In this section we study full non-perturbative equations including non-longitudinal modes, {\em i.e.,} the modes other than $f$ (and $g$). Our goal is to study non-perturbative massive solutions. For massive solutions we can focus on field configurations with no spatial dependence. This can be thought about in two ways. In the rest frame of a massive object the spatial momenta vanish, so a solution to the (nonlinear) ``wave" equation depends only on time.\footnote{{\em E.g.}, in the case of massive Klein-Gordon equation $\left(\partial^M\partial_M - m^2\right)\phi = 0$, in the rest frame the solution is given by $\phi = a~\cos(Et) + b~\sin(Et)$, where the energy $E = m$.} Alternatively, as in Subsection 4.2, we can compactify the spatial coordinates on a torus $T^{D-1}$ and disregard the Kaluza-Klein modes. This is a valid approximation for slow-moving massive objects.

{}We will therefore look for solutions of the following form, which depend only on time $t$:
\begin{eqnarray}
 &&G_{00} \equiv -u(t)~,\\
 &&G_{ij} \equiv G_{ij}(t)~,\\
 &&G_{i0} \equiv 0~.
\end{eqnarray}
While $u(t) > 0$ in the metric can be absorbed into the definition of time via $d\tau = u^{1/2} dt$, we must keep it for the purpose of solving the equations of motion because diffeomorphisms are broken by the potential $V$ and $u$ contributes into $V$.

{}For the above field configurations we have:
\begin{eqnarray}\label{Rij}
 && R_{ij} = {1\over 2u}~G_{ij}^{\prime\prime} + {1\over 4u} \left(\phi - {u^\prime\over u}\right)~G_{ij}^\prime - {1\over 2u}~f_{ij}~,\\
 && R_{00} = -{1\over 2}~\phi^\prime + {u^\prime\over 4u}~\phi - {1\over 4}~G^{ij}~f_{ij}~,\\
 && R_{i0} = 0~,
\end{eqnarray}
where
\begin{eqnarray}
 &&\phi \equiv G^{ij}~G_{ij}^\prime~,\\
 &&f_{ij} \equiv G_{ik}^\prime~G_{jl}^\prime~G^{kl}~.
\end{eqnarray}
The equations of motions read:
\begin{eqnarray}\label{R-V}
 &&R_{MN} = \mu^2 \left[V^\prime(X)~\eta_{MN} +  {{V(X) - X~V^\prime(X)}\over {D-2}}~G_{MN}\right]~,\\
 &&\partial_M\left[\sqrt{-G}V^\prime(X) G^{MN}\right] = 0~,\label{Bianchi1}
\end{eqnarray}
where
\begin{equation}
 X\equiv u^{-1} + G^{ij}\eta_{ij}~.
\end{equation}
When $u\equiv 1$ and $G_{ij} = \eta_{ij}$ (the Minkowski background), we have $X = D$.

{}From the Bianchi identity (\ref{Bianchi1}) we have
\begin{equation}\label{u}
 u = {1\over \nu^2}\det(G_{ij})\left[V^\prime(X)\right]^2~,
\end{equation}
where $\nu$ is an integration constant.

{}In the general case solving the above nonlinear equations analytically is challenging (albeit not impossible -- see Subsection 7.3 below). For the linear potential $V(X) = X - (D-2)$ we have substantial simplifications. First, from (\ref{u}) we have
\begin{eqnarray}
 && u = {1\over \nu^2}\det(G_{ij})~,\\
 && u^\prime = \phi~u~,
\end{eqnarray}
and (\ref{Rij}) simplifies as follows:
\begin{equation}
 R_{ij} = {1\over 2 u}\left[G_{ij}^{\prime\prime} - f_{ij}\right]~.
\end{equation}
So the equations of motion read:
\begin{eqnarray}
 &&G_{ij}^{\prime\prime} - f_{ij} = 2\mu^2 u\left[\eta_{ij} - G_{ij}\right]~,\\
 &&\phi^\prime - {1\over 2}~\phi^2 + {1\over 2}~G^{ij} f_{ij} = 2\mu^2 [1 - u]~.
\end{eqnarray}
From the $G_{ij}$ equation we also have
\begin{equation}\label{phiprime}
 \phi^\prime = 2\mu^2 u\left[G^{ij}\eta_{ij} - (D - 1)\right]~.
\end{equation}
We will now analyze the above equations.

\subsection{Three Dimensions}

{}In $D = 3$ we have the following simplification:
\begin{eqnarray}
 &&f_{ij} = \phi~G_{ij}^\prime - \xi~G_{ij}~,\\
 &&\xi \equiv {\det(G_{ij}^\prime)\over\det(G_{ij})}~,\label{xi}
\end{eqnarray}
so we have
\begin{eqnarray}\label{Gij}
 &&G_{ij}^{\prime\prime} - \phi~G_{ij}^\prime + \xi~G_{ij} = 2\mu^2 u\left[\eta_{ij} - G_{ij}\right]~,\\
 &&\phi^\prime - \xi = 2\mu^2 [1 - u]~.
\end{eqnarray}
Using (\ref{phiprime}) we also have
\begin{equation}
 \xi = 2\mu^2 \left[u(G^{ij}\eta_{ij} - 1) - 1\right]~.
\end{equation}
Let
\begin{equation}
 \chi \equiv \eta^{ij}G_{ij}~.
\end{equation}
If we assume that $\chi = 2$, then from (\ref{Gij}) it follows that $\xi = 0$ and $\det(G_{ij}^\prime) = 0$. However, in $D = 3$ this would imply that $G_{ij} = \eta_{ij}$. Indeed, if $\chi = 2$, then we have $G_{ij} = \eta_{ij} + a~(\sigma_3)_{ij} + b~(\sigma_1)_{ij}$, where $\sigma_1$ and $\sigma_3$ are the Pauli matrices, and $\det(G_{ij}^\prime) = -(a^\prime)^2 - (b^\prime)^2$, so $\xi = 0$ implies that $a$ and $b$ are constant (and can be set to 1 by rescaling the spatial coordinates). This means that for the linear potential in $D=3$ nontrivial solutions always involve a mixture of the longitudinal and non-longitudinal modes.

{}To summarize, our equations of motion read:
\begin{eqnarray}
 && G_{ij}^{\prime\prime} - \phi~G_{ij}^\prime + \xi~G_{ij} = 2\mu^2 u\left[\eta_{ij} - G_{ij}\right]~,\\
 && \xi = 2\mu^2 \left[u(G^{ij}\eta_{ij} - 1) - 1\right]~,\\
 && u = {1\over \nu^2}\det(G_{ij})~,\\
 && \phi = {u^\prime \over u}~.
\end{eqnarray}
We have the following solution:
\begin{eqnarray}
 &&G_{11} = u~,\\
 &&G_{22} = 1~,\\
 &&G_{12} = 0~,\\
 &&\nu = 1~,
\end{eqnarray}
and $u$ is a solution to the following equation:
\begin{equation}
 \left({u^\prime \over u}\right)^\prime = 2\mu^2 [1 - u]~.
\end{equation}
Let
\begin{equation}
 u \equiv e^\omega~.
\end{equation}
Then we have
\begin{equation}
 \omega^{\prime\prime} = 2 \mu^2\left[1 - e^\omega\right]~,
\end{equation}
and
\begin{equation}\label{omega}
 (\omega^\prime)^2 + \Phi(\omega) = 4\mu^2\eta~,
\end{equation}
where $\eta$ is an integration constant,
\begin{equation}
 \Phi(\omega)\equiv 4\mu^2\left[e^\omega - 1 - \omega\right]~,
\end{equation}
and we have $\Phi(\omega)\geq 0$ and $\eta\geq 0$. For $\eta = 0$ we have a trivial solution $\omega \equiv 0$ (and $u \equiv 1$). For $\eta > 0$ we have solutions oscillating between $\omega_-$ and $\omega_+$, where $\omega_-$ and $\omega_+$ are the negative and positive roots of the equation
\begin{equation}\label{omegapm}
 e^{\omega_\pm} - 1 - \omega_\pm = \eta~,
\end{equation}
and the period of these oscillations is given by:
\begin{equation}\label{T}
 T = {1\over \mu}~F(\eta)~,
\end{equation}
where
\begin{equation}
 F(\eta) \equiv \int_{\omega_-}^{\omega_+} {d\omega\over\sqrt{\eta + 1 + \omega - e^\omega}}~.
\end{equation}
For $\eta \ll 1$ we have $T\approx 2\pi/(\sqrt{2}\mu)$, which is consistent with fact that in this case perturbatively the mass of both the longitudinal (see Subsection 6.1) and of the non-longitudinal (see (\ref{zeta})) modes is $\sqrt{2}\mu$. However, if $\eta$ is not small, then the oscillation period depends on the amplitude, which is controlled by $\eta$. Furthermore, non-perturbatively we have no ghost.

{}The above oscillating solutions are interesting because only one of the spatial directions expands and contracts, while the other does not. Such unisotropically oscillating solutions may have interesting applications in the context of cosmology.

{}Finally, let us note that the above solution can be rotated, and the following is also a solution to the equations of motion:
\begin{eqnarray}
 &&G_{11} = u~\cos^2(\alpha) + \sin^2(\alpha)~,\\
 &&G_{22} = u~\sin^2(\alpha) + \cos^2(\alpha),\\
 &&G_{12} = (u - 1)\cos(\alpha)\sin(\alpha)~,\\
 &&\nu = 1~,
\end{eqnarray}
where $u$ is the same as above and $\alpha$ is a constant.

\subsection{``Cosmological Strings"}

{}Looking at the equations of motion for the linear potential for general $D$, it is evident that solutions with only one oscillating spatial direction exist in any $D$. Up to $SO(D-1)$ rotations, these solutions are given by:
\begin{eqnarray}
 &&G_{11} = u \equiv e^\omega~,\\
 &&G_{ii} = 1,~~~i > 1\\
 &&G_{ij} = 0~,~~~i\not=j\\
 &&\nu = 1~,
\end{eqnarray}
and the metric is given by
\begin{equation}\label{string}
 ds^2 = e^{\omega(t)}~\left[-dt^2 + (dx^1)^2\right] + \sum_{i=2}^{D-1} (dx^i)^2~,
\end{equation}
where $\omega$ is an oscillating solution to the equation (\ref{omega}) with the period $T$ given by (\ref{T}). In terms of the time coordinate $\tau$ defined via
\begin{equation}
 d\tau \equiv e^{\omega/2}~dt~,
\end{equation}
we have the metric
\begin{equation}
 ds^2 = -d\tau^2 + e^\omega(dx^1)^2 + \sum_{i=2}^{D-1} (dx^i)^2~.
\end{equation}
The oscillation period in terms of the time coordinate $\tau$ is given by
\begin{equation}
 {\widetilde T} = {1\over \mu}~{\widetilde F}(\eta)~,
\end{equation}
where
\begin{equation}
 {\widetilde F}(\eta) \equiv \int_{\omega_-}^{\omega_+} {e^{\omega/2}d\omega\over\sqrt{\eta + 1 + \omega - e^\omega}}~,
\end{equation}
and $\omega_\pm$ are defined in (\ref{omegapm}).

{}These solutions are 2-dimensional cosmological defects -- ``cosmological strings" -- in a $D$-dimensional space-time. Unlike static cosmic strings, cosmological strings are not static objects. The cosmological sting solutions we found here have the length scale (``warp" factor) in one spatial dimension oscillating in time, while the other $(D-2)$ spatial dimensions remain static and flat.

{}Cosmological strings arise in the gravitational Higgs mechanism because diffeomorphisms are spontaneously broken. Indeed, for the metric of the form (\ref{string}) and, more generally, for any metric of the form
\begin{equation}
 ds^2 = \gamma_{\mu\nu}dx^\mu dx^\nu + \sum_{i=2}^{D-1} (dx^i)^2~,
\end{equation}
where the metric $\gamma_{\mu\nu}$, $\mu,\nu = 0,1$ is independent of $x^i$, $i>1$, we have
\begin{eqnarray}
 &&R_{\mu\nu} = {1\over 2}\gamma_{\mu\nu} R~,\\
 &&R_{ij} = 0~,~~~i,j>1~,\\
 &&R_{\mu i} = 0,~~~i>1~,
\end{eqnarray}
and the Einstein tensor $E_{MN}\equiv R_{MN} - {1\over 2}G_{MN} R$ is given by:
\begin{eqnarray}
 &&E_{\mu\nu} = 0~,\\
 &&E_{ij} = -{1\over 2}\eta_{ij} R~,~~~i,j>1~,\\
 &&E_{\mu i} = 0~,~~~i>1.
\end{eqnarray}
So, to have such solutions, we must have non-vanishing $E_{ij}$ in the transverse directions $i,j>1$. This is precisely what transpires in the gravitational Higgs mechanism with the linear potential $V(X) = X - (D-2)$. In fact, it is not difficult to see that such solutions exist for no other potential $V(X)$ (albeit {\em a priori} this does not exclude more general cases -- see footnote \ref{foot} hereof).

\subsection{Constant-volume Solutions}

{}Above we discussed non-perturbative solutions for the linear potential. In this subsection we discuss purely non-longitudinal solutions, which we will also refer to as ``traceless" solutions, for which $\eta^{ij}G_{ij} = D-1$, {\em i.e.}, $h_{ij}\equiv G_{ij} - \eta_{ij}$ is traceless: $h\equiv \eta^{ij} h_{ij} = 0$. For the general potential $V(X)$ we can achieve a simplification by working in $D=3$ and considering constant-volume solutions for which
\begin{equation}\label{detG}
 \left[\det(G_{ij})\right]^\prime = 0~.
\end{equation}
In this case the equations of motion simplify as follows:
\begin{eqnarray}
 &&G^{\prime\prime}_{ij} + \xi~G_{ij} = 2\mu^2~u~\left(V^\prime(X)~\eta_{ij} + \left[V(X) - X~V^\prime(X)\right]~G_{ij}\right)~,\\
 &&\xi = -2\mu^2~\left(V^\prime(X) + \left[V(X) - X~V^\prime(X)\right]~u\right)~,\label{xi-eom}
\end{eqnarray}
where the first equation follows from the $ij$ component of the equations of motion (\ref{R-V}), while the second equation follows from the $00$ component, and
\begin{eqnarray}
 &&\xi\equiv {{\det(G^\prime_{ij})}\over{\det(G_{ij})}}~,\\
 &&X\equiv u^{-1} + G^{ij}\eta_{ij}~.
\end{eqnarray}
The $00$ component of the metric, $u\equiv -G_{00}$, is determined via (\ref{u}), which follows from the Bianchi identity (\ref{Bianchi1}):
\begin{equation}\label{u1}
 u = {1\over \nu^2}\det(G_{ij})\left[V^\prime(X)\right]^2~,
\end{equation}
where $\nu$ is an integration constant. Note that in $D=3$, when $\eta^{ij}h_{ij} = 0$, we have
\begin{equation}
 G^{ij} = \left[\eta^{ij} - h^{ij}\right]~\left[\det(G_{ij})\right]^{-1}~,
\end{equation}
where $h^{ij}\equiv\eta^{ik}\eta^{jl}h_{kl}$ and $\eta_{ij}h^{ij} = 0$. This implies that $G^{ij}\eta_{ij} = 2\left[\det(G_{ij})\right]^{-1}$ is constant as we have (\ref{detG}), so both $X$ and $u$ are also constant. Then from (\ref{xi-eom}) it follows that $\xi$ is also constant. We therefore have:
\begin{eqnarray}
 &&h_{ij}^{\prime\prime} = - {\widetilde M}^2~h_{ij}~,\\
 &&(1+u)~V^\prime(X) + 2u~\left[V(X) - X~V^\prime(X)\right] = 0~,\label{u-X}\\
 &&\xi=-\mu^2(1-u)V^\prime(X)~,\\
 &&X = u^{-1} + 2~\left[\det(G_{ij})\right]^{-1}~,\\
 &&{\widetilde M}^2 \equiv 2\mu^2 u V^\prime(X)~.
\end{eqnarray}
A solution is given by
\begin{eqnarray}
 &&h_{11} = -h_{22} = \rho~\cos({\widetilde M}(t-t_0))~,\\
 &&h_{12} = \rho~\sin({\widetilde M}(t-t_0))~,\\
 &&u = {{1-\rho^2}\over{1+\rho^2}}~,
\end{eqnarray}
with
\begin{eqnarray}
 && X = {{3+\rho^2}\over{1-\rho^2}}~,\\
 &&\det(G_{ij}) = 1-\rho^2~,\\
 &&\xi = -{\rho^2~\over{1-\rho^2}}~{\widetilde M}^2,
\end{eqnarray}
and $\rho$ is subject to (\ref{u-X}). Using $\rho^2 = (X-3)/(X+1)$ and $u=2/(X-1)$, (\ref{u-X}) reduces to
\begin{equation}\label{X}
 4V(X) - (3X - 1)V^\prime(X) = 0~.
\end{equation}
Note that due to (\ref{cosm.const}) this equation always has at least one solution, namely, $X=3$ (so $u=1$ and $\rho=0$), which is simply the Minkowski background. Also note that, in agreement with our results in Subsection 7.1, for the linear potential $V(X) = X - 1$, this is the only solution. However, for nonlinear potentials we can have other nontrivial solutions to (\ref{X}). Such a solution $X = X_*$ must satisfy two conditions: $X_* > 3$ (so that $\rho^2$ is positive -- in this case $u$ is also positive) and $V^\prime(X_*) > 0$ (so that ${\widetilde M}^2$ is positive).

{}We have such a solution already for a quadratic potential:
\begin{equation}\label{VX*}
 V(X) = \lambda\left[X^2 + 2(X_* + 2)X - (2X_* + 1)\right]~.
\end{equation}
For this potential (\ref{X}) has two solutions: $X=3$ and $X=X_*$. We can choose $X_* > 3$. We also have $V^\prime(X_*) = 4\lambda(X_* + 1)$, which is positive for $\lambda > 0$. If we compute perturbative mass squared and $\zeta$ via (\ref{M}) and (\ref{zeta}), we get $M^2 = 2\mu^2\lambda (X_* + 3) > 0$ and $\zeta = 2(X_* + 5)/(X_* + 3)$, so perturbatively the traceless components have positive mass squared, and so does the trace component (as $1 < \zeta < 3$ -- see footnote \ref{foot2} hereof), albeit perturbatively the trace component is a propagating ghost (this is a non-Fierz-Pauli point), while non-perturbatively there is no ghost as the Hamiltonian (\ref{Hamiltonian}) is evidently positive-definite in this case.

{}Let us rescale the time coordinate via
\begin{equation}
 d\tau^2 \equiv u~dt^2
\end{equation}
Then the metric reads
\begin{eqnarray}
 &&ds^2 = -d\tau^2 + \left[1 + \rho~\cos(M_1\tau)\right](dx^1)^2 + \left[1 - \rho~\cos(M_1\tau)\right](dx^2)^2 + \nonumber\\
 &&\,\,\,2\rho~\sin(M_1\tau)~(dx^1)(dx^2)~,
\end{eqnarray}
where $M_1^2 \equiv {\widetilde M}^2 / u = 2\mu^2 V^\prime(X_*)$.

{}Finally, let us mention that in the above constant-volume traceless solutions the amplitude $\rho$ of the oscillations is determined by the higher-derivative couplings in the scalar sector. Indeed, $\rho$ is fixed by $X_*$, which in turn is fixed via the higher-derivative coupling once we normalize the kinetic term, {\em i.e.}, the term linear in $X$ in (\ref{VX*}), by setting $2\lambda(X_*+2) = 1$. Also note that $\rho$ is small when $X_* - 3$ is small.

\section{An Aside: New Massive Gravity}

{}In this section, as an illustrative aside, we discuss a simple application of the methods discussed in the previous section to New Massive Gravity \cite{BHT}, which appears to confirm the results of \cite{dRGPTY} obtained via a different framework. For computational convenience, our discussion here will be for general dimension $D$ and a general higher derivative gravity action (without any scalars or breaking of the diffeomorphism invariance) subject to the restrictions on higher curvature terms that (\ref{sumk}) does not contain higher derivative terms (see Section 5). We will then apply it to the particular case of New Massive Gravity.

{}For the field configurations of the form
\begin{equation}
 G^{MN} = {\rm diag}(g(t)~\eta^{00}, f(t)~\eta^{ii})~,
\end{equation}
the dimensionally reduced action reads
\begin{eqnarray}\label{NMG}
 S = -\kappa \int dt ~g^{-{1\over 2}} e^{-(D-1)q}~Q(gU^2)~,
\end{eqnarray}
where
\begin{equation}
 U\equiv \partial_t q \equiv {1\over 2}\partial_t \ln(f)
\end{equation}
and $Q(gU^2)$ is given by (\ref{sumk}). The $g$ equation of motion reads:
\begin{equation}\label{g.eom.nmg}
 2gU^2Q^\prime(gU^2) = Q(gU^2)~.
\end{equation}
The Hamiltonian ${\cal H} = 0$, which is due to the fact that diffeomorphisms are unbroken. Due to (\ref{g.eom.nmg}), the $q$ equation of motion reduces to
\begin{equation}
 \partial_t\left[g^{1\over 2}UQ^\prime(gU^2)\right] = 0~.
\end{equation}
Let us introduce the time coordinate $\tau$ via
\begin{equation}
 d\tau = g^{-{1\over 2}} dt~.
\end{equation}
The metric reads
\begin{equation}
 ds^2 = -d\tau^2 + e^{-2q}~\eta_{ij}dx^idx^j~.
\end{equation}
We then have
\begin{equation}
 \left(\partial_\tau q\right)^2 = x_*~,
\end{equation}
where $x_*$ is a solution to the equation
\begin{equation}\label{vacua}
 2x_*Q(x_*) = Q(x_*)~,
\end{equation}
and
\begin{equation}
 Q(x) \equiv \sum_{k=1}^\infty c_k x^k~.
\end{equation}
So, we have a Minkowski solution as $x_* = 0$ is a solution to (\ref{vacua}). Depending on $Q$, there might also exist other solutions. Consider the case of quadratic $Q$:
\begin{equation}
 Q(x) = c_1 x + c_2 x^2~.
\end{equation}
Then we have an additional solution
\begin{equation}
 x_* = -{c_1\over 3c_2}~.
\end{equation}
For the action
\begin{equation}
S = M_P^{D-2}\int d^Dx \sqrt{-G}\left[\sigma R + \alpha R^2 + \beta R^{MN} R_{MN}\right]~,
\end{equation}
where
\begin{eqnarray}
 &&\beta \equiv {1\over m^2}~,\\
 &&\sigma = \pm 1~,
\end{eqnarray}
we have (see Section 5)
\begin{eqnarray}
 &&\alpha = -{D\over 4(D-1)m^2}~,\\
 &&c_1 = \sigma(D-1)(D-2)~,\\
 &&c_2 = {(D-1)(D-2)^2(4-D)\over 12 m^2}~,\\
 &&x_* = -{4\sigma m^2\over (4-D)(D-2)}~,
\end{eqnarray}
where we are assuming $D\not =4$.\footnote{For $D=4$ we have $c_2 = 0$ and there is only $x_* = 0$ solution.}
For $\sigma=-1$ we have de Sitter background $q = H\tau$ with the Hubble parameter $H$ and the cosmological constant ${\widetilde \Lambda}$:
\begin{eqnarray}
 &&H^2 = {4m^2\over(4-D)(D-2)}~,\\
 &&{\widetilde \Lambda} = (D-1)(D-2)H^2 = 4(D-1)m^2/(4-D)~.
\end{eqnarray}
In $D=3$ this reproduces the known results in New Massive Gravity. Our point here is that, with the restrictions of Section 5, only first time-derivatives appear in action (\ref{NMG}) and no trouble is expected with the longitudinal mode. This is evident in the language of $g$ and $f$, and appears to be much more nontrivial in the usual perturbative parametrization (see \cite{dRGPTY}). In particular, non-perturbatively there is no unusual dynamics associated with the longitudinal mode, which simply determines the choice of the background (Minkowski or de Sitter). Therefore, there appears to be no issue with unitarity in New Massive Gravity. What is less evident is whether there exist non-perturbative massive solutions for the non-longitudinal modes as they do in the gravitational Higgs mechanism (see Section 7). In this regard, it would be interesting to apply the methods discussed in Section 7 to New Massive Gravity and see if there exist non-perturbative massive solutions to the (highly nonlinear) equations of motions. We will do this in the remainder of this section.

\subsection{No Cosmological Strings in New Massive Gravity}

{}In the gravitational Higgs mechanism, in the previous section, we found cosmological string solutions. Here we show that such solutions are absent in New Massive Gravity. As before, for computational convenience we will work in the general dimension $D$. We start with the action (\ref{NMG}). The equations of motion read:
\begin{equation}\label{EK}
 \sigma E_{MN} + K_{MN} = 0~,
\end{equation}
where
\begin{eqnarray}
 && E_{MN} \equiv R_{MN} - {1\over 2} G_{MN} R~,\label{E}\\
 &&K_{MN} \equiv (2\alpha +\beta)(G_{MN}\nabla^2 - \nabla_M\nabla_N)R + \beta\nabla^2 E_{MN} + \nonumber\\
 &&\,\,\,2\alpha R \left(R_{MN} - {1\over 4}G_{MN} R\right) + 2\beta\left(R_{MSNT} - {1\over 4}G_{MN}R_{ST}\right)R^{ST}~.\label{K}
\end{eqnarray}
For any metric of the form
\begin{equation}
 ds^2 = \gamma_{\mu\nu}dx^\mu dx^\nu + \sum_{i=2}^{D-1} (dx^i)^2~,
\end{equation}
where the metric $\gamma_{\mu\nu}$, $\mu,\nu = 0,1$ is independent of $x^i$, $i>1$, we have
\begin{eqnarray}
 && E_{\mu\nu} = 0~,\\
 && E_{ij} = -{1\over 2}\eta_{ij}R~,~~~i,j > 1\\
 && K_{\mu\nu} = (2\alpha+\beta)\left[(\gamma_{\mu\nu}\nabla^2 - \nabla_\mu\nabla_\nu)R + {1\over 4}\gamma_{\mu\nu}R^2\right]~,\\
 && K_{ij} = \left[\left(2\alpha + {1\over 2}\beta\right)\nabla^2 R - {1\over 4}(2\alpha + \beta)R^2\right]\eta_{ij}~, ~~~i,j>1.
\end{eqnarray}
The equations of motion (\ref{EK}) then imply that
\begin{eqnarray}
 &&\nabla_\mu\nabla_\nu R = -{1\over 4}\gamma_{\mu\nu} R^2~,\\
 &&(3\alpha + \beta)R^2 + \sigma R = 0~,
\end{eqnarray}
which only have a trivial solution $R = 0$ assuming $\sigma \not=0$. For $\sigma = 0$ (no Einstein-Hilbert term) we can satisfy the second equation if the condition $3\alpha + \beta = 0$ is satisfied.\footnote{This condition is not satisfied in New Massive Gravity in $D=3$. This condition is the same as (\ref{alphabeta}) in $D=4$, in which case we simply have Weyl gravity up to the Gauss-Bonnet combination, which is a total derivative in $D=4$.} However, these cannot be massive solutions. This can be seen by transforming the 2-dimensional metric $\gamma_{\mu\nu}$ into a conformally flat form. Alternatively, if we look for solutions that depend only on time $t$ (see the discussion at the beginning of Section 7), then we invariably have $R=0$. So, there are no cosmological strings in New Massive Gravity.\footnote{In the gravitational Higgs mechanism we found cosmological string solutions for the linear potential.}

\subsection{Analysis of Massive Solutions}

{}In this subsection we study full non-perturbative equations in New Massive Gravity in the context of non-perturbative massive solutions. The theory has full diffeomorphism invariance. Therefore, we can always set $G_{00} = 1$ and $G_{i0} = 0$. For the remaining metric components we look for solutions that depend only on time $t$. We then have
\begin{eqnarray}
 &&R_{00} = -{1\over 2} R + z~,\\
 &&G^{ij}~R_{ij} = {1\over 2} R + z~,\label{RG}\\
 &&R_{i0} = 0~.
\end{eqnarray}
Using our formulas in Section 7, in $D=3$, which we focus on here, we have
\begin{equation}
 z = {\xi\over 4}~,
\end{equation}
where $\xi$ is defined in (\ref{xi}).

{}In $D=3$ we have the following identity:
\begin{equation}\label{RGR}
 R_{ik}R_{jl}G^{kl} = R_{ij}\left(R_{kl}G^{kl}\right) - G_{ij}~{\det(R_{kl})\over\det(G_{kl})}~.
\end{equation}
We are assuming that $\det(G_{kl})>0$. It then follows that
\begin{eqnarray}\label{RMNsq}
 R_{MN}R^{MN} = {1\over 2}~R^2 + 2~z^2 - 2~{\det(R_{kl})\over\det(G_{kl})}~.
\end{eqnarray}
On the other hand, the equation of motion (\ref{EK}) in $D=3$ implies that
\begin{equation}\label{traceK}
 R_{MN}R^{MN} - {3\over 8}~R^2 = -m^2~R~,
\end{equation}
where we have set $\beta = 1/m^2$ and $\sigma=-1$. Furthermore,
\begin{eqnarray}\label{EOMEK}
 &&E_{MN} = K_{MN}~,\\
 &&m^2~K_{MN} = {1\over 4}\left(G_{MN}\nabla^2 - \nabla_M\nabla_N\right)R + \nabla^2 E_{MN} + \nonumber\\
 &&\,\,\, {9\over 4}R_{MN}R - 4R_{MS}{R^S}_N -{3m^2\over 2}G_{MN} R - {1\over 4}G_{MN}R^2~.
\end{eqnarray}
From (\ref{RMNsq}) and (\ref{traceK}) it follows that
\begin{equation}\label{R1}
 R ={2\over m^2}\left[{\det(R_{kl})\over\det(G_{kl})} - z^2 - {1\over 16}~R^2\right]~.
\end{equation}
Furthermore, from the $00$ component of the equation of motion (\ref{EOMEK}), we also have:
\begin{equation}\label{R2}
 R = {2\over 3m^2} \left[z^{\prime\prime} + \left(m^2 +{7\over 4}~R\right)~z - 4~z^2 - {1\over 8}~R^2\right]~,
\end{equation}
where we have taken into account that $E_{00} = R_{00} + R/2 = z$ and that $\nabla_0^2 E_{00} = E_{00}^{\prime\prime}$. Finally, the $ij$ components of the equation of motion (\ref{EOMEK}) read
\begin{equation}\label{Rij1}
 -\nabla_0^2 R_{ij} + \left({1\over 4}R - 4z - m^2\right)R_{ij} + \left(m^2 R + 4z^2 + {1\over 4}R^{\prime\prime}\right)G_{ij} = 0~,
\end{equation}
where we have used (\ref{RGR}), (\ref{RG}) and (\ref{R1}). Let
\begin{equation}
 {\overline R}_{ij} \equiv R_{ij} - {1\over 2}G_{ij}\left(G^{kl}R_{kl}\right) = R_{ij} - G_{ij}\left({R\over 4} + {z\over 2}\right)~.
\end{equation}
Note that $G^{ij}{\overline R}_{ij} = 0$. Using (\ref{RG}) and the fact that $\nabla_0^2 z = z^{\prime\prime}$, from (\ref{Rij1}) we get
\begin{equation}\label{overlineR}
 -\nabla_0^2 {\overline R}_{ij} + \left({1\over 4}R - 4z - m^2\right){\overline R}_{ij} = 0~.
\end{equation}
Also,
\begin{equation}
\det(R_{ij}) = \det({\overline R}_{ij}) + \left({R\over 4} + {z\over 2}\right)^2~\det(G_{ij})~,
\end{equation}
so (\ref{R1}) reads
\begin{equation}
 R = {2\over m^2}\left[{\det({\overline R}_{kl})\over\det(G_{kl})} - {3\over 4}~z^2 - {1\over 4}~R~z\right]~.
\end{equation}
Note that $G_{ij} = e^{-2Ht}\eta_{ij}$ is a solution to (\ref{R1}), (\ref{R2}) and (\ref{Rij1}) for $H^2=4m^2$. Indeed, in this case we have $R_{ij} = 2H^2 G_{ij}$, $R_{00} = -2H^2$, $z = H^2$ and $R=6H^2$. This is the de Sitter solution discussed above. However, what we are interested in here is finding massive solutions. Just as in the case of gravitational Higgs mechanism, we expect that if such solutions exist, they should exist for weak-field configurations.

\subsubsection{Linearized Approximation}

{}In the linearized approximation
\begin{equation}
 G_{ij} \equiv \eta_{ij} + h_{ij}~,
\end{equation}
and one keeps only the terms linear in $h_{ij}$ in the equations of motion. Thus, we have
\begin{eqnarray}
 &&R_{ij}^{(1)} = {1\over 2}~h_{ij}^{\prime\prime}~,\\
 &&R_{00}^{(1)} = -{1\over 2}~h^{\prime\prime}~,\\
 &&R_{i0}=0~,\\
 &&R^{(1)} = h^{\prime\prime}~,\label{h}
\end{eqnarray}
where $h\equiv \eta^{ij}h_{ij}$. Note that the leading term in $z$ is quadratic in $h_{ij}$, so $z^{(1)} = 0$. Both (\ref{R1}) and (\ref{R2}) then require that
\begin{equation}
 R^{(1)} = 0~,
\end{equation}
and at this order the trace component vanishes: $h^{(1)} = 0$. We have
\begin{eqnarray}
 &&K_{MN}^{(1)} = -{1\over m^2}~\partial_t^2 E^{(1)}_{MN}~,\\
 &&E^{(1)}_{MN} = R^{(1)}_{MN}~,
\end{eqnarray}
so the equations of motion (\ref{EK}) give
\begin{equation}
 \left[\partial_t^2 + m^2\right]~\partial_t^2 h_{ij}^{(1)} = 0~,
\end{equation}
which are solved by solutions to the equations
\begin{equation}\label{hij1}
 \left[\partial_t^2 + m^2\right]~h_{ij}^{(1)} = 0~,
\end{equation}
which are massive oscillating solutions.

\subsubsection{Next-to-linear (Quadratic) Order}

{}Let $h_{ij}^{(1)}$ be an arbitrary nontrivial solution of (\ref{hij1}). We have $h^{(1)} = 0$ and
\begin{equation}\label{z1}
 z^{(2)} = {1\over 4}~\det(\partial_t h^{(1)}_{ij})~.
\end{equation}
Using (\ref{hij1}) we then have
\begin{eqnarray}\label{z2}
 && z^{(2)} = C - {m^2\over 4}~\det(h_{ij}^{(1)})~,\\
 && \partial_t^2 z^{(2)} = {m^4\over 2}~\det(h_{ij}^{(1)}) - 2m^2~z^{(2)} = m^4~\det(h_{ij}^{(1)}) - 2m^2~C~.
\end{eqnarray}
Here $C$ is an integration constant ($C^\prime \equiv 0$) -- (\ref{z2}) can be obtained by differentiating (\ref{z1}) w.r.t. $t$ and using (\ref{hij1}). Note that, unless $h^{(1)}_{ij} \equiv 0$, we have $\det(\partial_t h^{(1)}_{ij}) < 0$ and $\det(h^{(1)}_{ij}) < 0$ as $h^{(1)}=0$, which implies that $C < 0$. Conversely, if $C=0$, then $h^{(1)}_{ij} \equiv 0$.

{}Next, we have
\begin{equation}
 \det(R^{(1)}_{ij}) = {m^4\over 4}~\det(h^{(1)}_{ij})~.
\end{equation}
At the quadratic order in $h_{ij}$ and its derivatives, (\ref{R1}) and (\ref{R2}) then give, respectively:
\begin{eqnarray}\label{R(2)}
 && R^{(2)} = {m^2\over 2}~\det(h^{(1)}_{ij})~,\\
 && R^{(2)} = {m^2\over 2}~\det(h^{(1)}_{ij}) - {2C\over 3}~,
\end{eqnarray}
which implies that $C = 0$ and $h^{(1)}_{ij} \equiv 0$. Thus, as we see, the quadratic order equations of motion have no massive oscillating solutions around the Minkowski background.

\subsubsection{Non-perturbative Argument}

{}The fact that perturbatively there are no massive solutions at the next-to-linear order indicates that either massive solutions do not exist or they cannot be treated perturbatively. Here we wish to see this without doing perturbative expansion. If there are massive solutions, we expect that they should be traceless, {\em i.e.}, in the decomposition $G_{ij} = \eta_{ij} + h_{ij}$, without assuming that $h_{ij}$ are small, we have $h\equiv\eta^{ij}h_{ij} = 0$. For such field configurations we have $\eta^{ij}G_{ij} = 2$ and
\begin{equation}
 \eta^{ij}{\overline R}_{ij} = 3z -{1\over 2}R~.
\end{equation}
From (\ref{overlineR}) we then have
\begin{equation}\label{R-second}
 R^{\prime\prime} = 8m^2 R - 16Rz + R^2 + 48 z^2~,
\end{equation}
where we have used (\ref{R2}). Note that this equation is exact for all traceless field configurations with $\eta^{ij}G_{ij} = 2$. It is now clear that there are no massive solutions corresponding to weak-field configurations. Indeed, for such field configurations $|R|$ and $|z|$ are small compared with $m^2$, so the last three terms on the r.h.s. of (\ref{R-second}) are small and can be neglected. However, the resulting equation $R^{\prime\prime} = 8m^2 R$ does not correspond to oscillating massive solutions; instead, it has exponentially growing solutions, which implies that the weak-field assumption is incorrect to begin with.\footnote{We can show directly that there are no constant-volume traceless solutions. Constant-volume solutions are defined as those with $\left[\det(G_{ij})\right]^\prime \equiv 0$. For such solutions we have $R = -2z$. Then (\ref{R2}) and (\ref{R-second}) imply that $z = 0$ or $z = 6m^2/25$. However, for traceless configurations $z < 0$ unless $h_{ij}\equiv 0$. This shows that there are no constant-volume traceless solutions.}

\subsection*{Acknowledgments}
{}I would like to thank Alberto Iglesias for reading the manuscript and valuable comments.


\end{document}